\documentstyle{jfm}
\RequirePackage{epsfig}

\newcommand{\beq}{\begin{equation}}
\newcommand{\eeq}{\end{equation}}
\newcommand{\beqarr}{\begin{eqnarray}}
\newcommand{\eeqarr}{\end{eqnarray}}
\newcommand{\barr}{\begin{array}}
\newcommand{\earr}{\end{array}}
\newcommand{\bcent}{\begin{center}}
\newcommand{\ecent}{\end{center}}
\newcommand{\rf}[1]{(\ref{#1})}

\renewcommand{\vec}[1]{\mbox{\boldmath $#1$}}
\newcommand{\vechat}[1]{{\skew3\hat{\vec{#1}}}}

\newcommand{\laplace}{\nabla^2}  

\newcommand{\cross}{\wedge}
\newcommand{\grad}{\bnabla}
\newcommand{\dvgnce}{\bnabla \cdot}
\newcommand{\curl}{\bnabla \wedge}

\newcommand{\pd}[1]{\upartial_{#1}}

\newcommand{\bess}{{\mathcal{B}}}
\newcommand{\ord}{{\mathcal{O}}}
\newcommand{\ex}{{\mathrm e}}
\newcommand{\im}{{\mathrm i}}

\newcommand{\uc}{\tilde{\vec{u}}}	
\newcommand{\ucr}{\tilde{u}_r}
\newcommand{\uct}{\tilde{u}_\theta}
\newcommand{\ucz}{\tilde{u}_z}
\newcommand{\vel}{\vec{u}}		
\newcommand{\magn}{\vec{B}}
\newcommand{\nlin}{\vec{N}}

\newcommand{\prandtl}{\xi}		
\newcommand{\hartmann}{Q}
\newcommand{\eqnlabel}[1]{\eqno{(\theequation {\it #1})}}
\newcommand{\eqntext}[1]{$$ \resetline #1 $$}

\newenvironment{eqnalph}
   {\refstepcounter{equation} \let\\=\eqnalpharr $$}
   {$$ \resetline}

\newcommand{\resetline}{\vspace{-\baselineskip}\newline}
\def\eqnalpharr[#1]{$$\vspace{-10pt}\vspace{#1}\resetline$$}

\catcode`\œ=\active \gdefœ{\setbox0=\hbox{0}\hbox to\wd0{}}
\def\Rey{\mbox{\it Re}}   
\ifCUPmtlplainloaded
\else
  \def\upi{\pi} 
  \def\upartial{\partial} 
\fi
\ifCUPmtlplainloaded

\fi
\ifCUPmtlplainloaded
  \font\bit = mtmib10 at 10.5pt \skewchar\bit ='177  
\else
  \font\bit = cmmib10 \skewchar\bit ='177  
\fi
\ifCUPmtlplainloaded
\else
  \font\tenbmi=cmmib10 at 10pt  \skewchar\tenbmi ='177
  \font\sevenbmi=cmmib10 at 7pt \skewchar\sevenbmi ='177
  \font\fivebmi=cmmib10 at 5pt  \skewchar\fivebmi ='177

  \newfam\bmifam
  \textfont\bmifam=\tenbmi
  \scriptfont\bmifam=\sevenbmi
  \scriptscriptfont\bmifam=\fivebmi
  
\fi
\newsavebox{\thalfbox}
\sbox{\thalfbox}{$\textstyle\frac{1}{2}$}

\newsavebox{\shalfbox}
\sbox{\shalfbox}{$\scriptstyle\frac{1}{2}$}

\newsavebox{\squartbox}
\sbox{\squartbox}{$\frac{1}{4}$} 

\newsavebox{\etbox}
\sbox{\etbox}{\boldmath$\eta$}

\newsavebox{\astrutbox}
\sbox{\astrutbox}{\rule[-5pt]{0pt}{20pt}}

\ifnfsstwo

\fi
\ifnfssone
  \newmathalphabet{\mathit}
    \addtoversion{normal}{\mathit}{cmr}{m}{it}
    \addtoversion{bold}{\mathit}{cmr}{bx}{it}

\fi
\ifoldfss

\fi

\mathchardef\varLambda="0103

\ifCUPmtlplainloaded
\else
\fi
\ifCUPmtlplainloaded
  \let\bcdot=\undefined
  \NewSymbolFont{bldsym}{mtbsy10}{'60}
  \NewMathSymbol{\bcdot}{2}{bldsym}{01}
\else
  \font\tenbms=cmbsy10          \skewchar\tenbms ='60
  \font\sevenbms=cmbsy10 at 7pt \skewchar\sevenbms ='60
  \font\fivebms=cmbsy10 at 5pt  \skewchar\fivebms ='60

  \newfam\bmsfam
  \textfont\bmsfam=\tenbms
  \scriptfont\bmsfam=\sevenbms
  \scriptscriptfont\bmsfam=\fivebms

  \edef\bsy{\hexnumber\bmsfam}
  \mathchardef\bnabla="0\bsy72
  \mathchardef\bcdotsymbol="0\bsy01
  \def\bcdot{\,\bcdotsymbol\,}
\fi
\def\etal{\mbox{\it et al.\ }}

\title[Hydromagnetic Taylor--Couette flow]
{Hydromagnetic Taylor--Couette flow.\\
Wavy modes.}

\author[A. P. Willis and C. F. Barenghi]%
{
   A.\ns P.\ns W\ls I\ls L\ls L\ls I\ls S\ns 
   \and 
   C.\ns F.\ns B\ls A\ls R\ls E\ls N\ls G\ls H\ls I\ls
}

\affiliation
{
   Department of Mathematics, University of Newcastle,\\
   Newcastle NE1 7RU, England
}

\pubyear{2002}
\volume{}
\pagerange{}
\date{March 2002}
\begin{document}

\maketitle

\begin{abstract}
   We investigate magnetic Taylor--Couette flow 
   in the presence of an imposed axial magnetic field.
   First we calculate nonlinear steady axisymmetric solutions and 
   determine how their strength depends on the applied magnetic field.
   Then we perturb these solutions to find the critical Reynolds
   numbers for the appearance of wavy modes, and the related wavespeeds, at
   increasing magnetic field strength.  
   We find that values of imposed magnetic field which alter only
   slightly the transition from circular--Couette flow to Taylor--vortex
   flow, can shift the transition from Taylor--vortex flow to wavy modes
   by a substantial amount.
   The results are compared against
   onset in the absence of a magnetic field.
\end{abstract}

\section{Introduction}
Taylor--Couette flow is one of the most important examples of fluid
systems which exhibits the spontaneous formation of increasingly complex
dynamic flow structures.  This occurs through a sequence of 
transitions which take place as the drive is increased (Andereck \etal 1986).
The first transition takes place when circular--Couette flow (CCF) becomes
unstable to axisymmetric disturbances which grow into a toroidal flow
pattern (Taylor--vortex flow --- TVF).  The second transition occurs
when the Taylor vortices become unstable to non-axisymmetric 
perturbations, which result in a variety of time-dependent flows 
(wavy modes).  More transitions take place at higher Reynolds numbers
and a great number of variations of the original pattern observed by 
\cite{taylor23} have been investigated
(Egbers \& Pfister 2000).

Our concern is Taylor--Couette flow in the presence of an axially 
imposed magnetic field.  Despite early theoretical (Chandrasekhar 1961)
and experimental interest (Donnelly \& Ozima 1962), this case has
been much less studied.  The success of recent dynamo experiments
(Gailitis \etal 2001; Steiglitz \& M\"uller 2001) 
and the astrophysical interest in the magneto-rotational instability
(Ji \etal 2001; R\"udiger \& Zhang 2001) add further motivations.

In our previous paper (Willis \& Barenghi 2002) we presented a 
convenient numerical formulation of the hydromagnetic Taylor--Couette
flow problem.  We determined critical Reynolds numbers for the appearance of 
axisymmetric Taylor vortices,  calculated finite amplitude solutions 
and compared our results with existing theoretical work and experiments.  
The aim of this paper
is to continue this investigation into the three-dimensional 
time-dependent flow regime of the wavy modes.

\section{Formulation and method of solution}
\label{sect:eqns}

Throughout this work we shall use the same notation as that of our previous
paper.  The fluid is contained between two concentric cylinders of
inner radius $R_1$ and outer radius $R_2$ which rotate at constant
angular velocity $\Omega_1$ and $\Omega_2$ respectively.  We make the
usual assumption that the cylinders have infinite height,
use cylindrical coordinates $(r,\theta,z)$, and assume that a constant
magnetic field $\magn_0=\mu_0H\vechat{z}$ is applied externally in the
axial direction.

We introduce the following dimensionless parameters: radius ratio ($\eta$),
Reynolds numbers ($\Rey_1$ and $\Rey_2$), Hartmann number ($Q$) and magnetic
Prandtl number ($\xi$) defined as
   \beq
      \eta = R_1/R_2,  \quad 
      \Rey_i = \frac{R_i \Omega_i \delta}{\nu},  \quad
      i=1,2, \quad
      \hartmann = \frac{\mu_0^2 H^2 \sigma \delta^2}{\rho \nu},  \qquad
      \prandtl = \frac{\nu}{\lambda} ,
   \eeq
where $\delta$ is the gap width, $\rho$ the density,
$\nu$ the kinematic viscosity, $\lambda$  the
magnetic diffusivity and $\mu_0$ the magnetic permeability.
Hereafter we assume that $\rho$, $\nu$, $\lambda$ and $\mu_0$ are constant.

The dimensionless equations governing incompressible hydromagnetic 
flow are
\begin{eqnalph}
   \label{eq:gov}
   \pd{t} \vel + (\vel \cdot \grad) \vel = 
   - \grad p + \laplace \vel 
   + \frac{Q}{\prandtl} (\curl \magn) \cross \magn,
   \qquad
   \dvgnce \vel = 0,
   \eqnlabel{a,b} 
   \\[0pt]
   \pd{t} \magn =
   \frac1{\prandtl} \laplace \magn + \curl (\vel \cross \magn),
   \qquad
   \dvgnce \magn = 0.
   \eqnlabel{c,d}
\end{eqnalph}
These equations have as a steady-state solution, $\uc$, the
circular--Couette flow, where
\beq
   \ucr = \ucz = 0, 
   \qquad
   \uct = Ar + B/r.
\eeq
The constants $A$ and $B$ are determined by the no-slip boundary 
conditions.  

Boundary conditions for the magnetic field can have a huge influence on
the flow, as seen by \cite{hollerbach01} when studying the analogous
problem in spherical geometry.  Here the axially imposed magnetic field
does not penetrate the boundaries but there are still differences 
between the theoretical results of \cite{chandrasekhar61} 
for insulating and perfectly conducting cylinders.  
However, in experiments by \cite{donnelly62} using mercury,
only a small difference was found between results
with Perspex and stainless--steel containers.
Therefore for practical reasons we take insulating
boundary conditions for the magnetic field.  
If the magnetic field is expanded over modes of the form
\beq
   \magn(r,\theta,z) = \magn(r) \, \ex^{\im (\alpha z + m\theta)},
\eeq
then the boundary conditions are
\beq
   \label{eq:mag_bcs}
   \left.
   \barr{rl}
   \alpha = m = 0:
   &
   B_\theta = B_z = 0;
   \\[2pt]
   \alpha=0, m\ne 0:
   &
   \pd{\theta} B_r = \pm m B_\theta,
   \quad
   B_z = 0 ;
   \\[2pt]
   \alpha \ne 0:
   &
   {\displaystyle
   \pd{z}B_r = \frac{\pd{r}\bess_m}{\bess_m} \, B_z ,
   \quad
   \frac{1}{r} \, \pd{\theta} B_z = \pd{z} B_\theta ,
   }
   \earr
   \right\}
\eeq
where for $\pm$ we take $+$ at $R_1$, $-$ at $R_2$.
The function $\bess_m(r)$ denotes the modified Bessel functions
$I_m(\alpha r)$ and $K_m(\alpha r)$, and the boundary condition 
\rf{eq:mag_bcs}
is evaluated at $R_1$ and $R_2$ respectively.

The magnetic Prandtl number, $\xi$, is very small in liquid metals
available in the laboratory
(mercury, $\xi=0.145\times 10^{-6}$; 
liquid sodium at 120$^\circ$C, $\xi=0.89\times 10^{-5}$).  
We set
   \beq
      \magn = \magn_0 + \prandtl \vec{b},
   \eeq
where $\magn_0$ is an externally applied field.
In the limit $\prandtl \to 0$ the equations governing the disturbance
to the circular--Couette flow become
\begin{eqnalph}
   \label{eq:gov_limit}
   (\pd{t} - \laplace) \vel' = \nlin - \grad p',
   \qquad
   \dvgnce \vel'  =  0,
   \eqnlabel{a,b}
   \\[-5pt]
   \laplace \vec{b}  =  \nlin_B,
   \qquad
   \dvgnce \vec{b} = 0,
   \eqnlabel{c,d}
\eqntext{where}
   \nlin  = 
   \hartmann (\curl \vec{b})\cross \magn_0
   - (\vel\cdot\grad) \vel' - (\vel'\cdot\grad) \uc,
   \qquad
   \nlin_B  = 
   - \curl (\vel \cross \magn_0) .
   \eqnlabel{e,f}
\end{eqnalph}
The finite amplitude disturbance $\vel'=\vel-\uc$ satisfies
homogeneous Dirichlet boundary conditions.
The magnetic field $\vec{b}$ satisfies the same boundary conditions as 
$\magn$ and is completely defined by the
velocity at any particular time.

The method of solution of (\ref{eq:gov_limit}{\it a-d})
was the topic of our earlier paper, \cite{willis02}, to which we refer
for further details, including tests of the numerical method and 
comparison with calculations performed by others.  Here it suffices to say
that to ensure divergence-free fields we adopt the toroidal-poloidal
decomposition
\beq
   \label{eq:pot_expansion}
   \vec{A} = \psi_0 \, \vechat{\theta} + \phi_0 \, \vechat{z}
   + \curl (\psi\vec{r}) + \curl \curl (\phi\vec{r}),
\eeq
where $\psi(r,t,z)$, $\phi(r,t,z)$ and $\psi_0(r)$, $\phi_0(r)$ 
contain the periodic and non-periodic parts of the field respectively.
The potentials $\psi$, $\phi$ are expanded over Fourier modes 
in the periodic coordinates and 
Chebyshev polynomials in the radial direction.

The governing equations for the magnetic field are the $r$-components
of the induction equation and its first curl.  Although it is 
commonplace to take the first and second curls for the velocity,
we follow the procedure applied to the magnetic field and take the
$r$-components of the momentum equation and its first curl.  
As the pressure has not been eliminated, we also take the divergence
to obtain the pressure--Poisson equation.

Whilst the decomposition \rf{eq:pot_expansion} raises the order of the
equations, the extra derivatives appear in the periodic coordinates,
so no extra boundary conditions are required.
All five governing equations are second order in $r$.  This property 
makes them easy to timestep stably and accurately, even for fully 
nonlinear three-dimensional flows, and simplifies implementation 
a great deal.  

The equations are collocated in the radial direction and 
the nonlinear terms (\ref{eq:gov_limit}{\it e,f})
are evaluated pseudo-spectrally where necessary.  
Explicit Adams--Bashforth timestepping is used on the nonlinear 
terms and implicit Crank--Nicolson on the linear terms.

\section{An imposed axial field}

The equation defining the 
magnetic field, (\ref{eq:gov_limit}{\it c}), may be re-written as
\beq
   \label{eq:curlcurlb}
   \curl\curl\vec{b} = 
   \curl
   ( \vel'\cross\vechat{z}),
\eeq 
since $\curl(\uc\cross\vechat{z})=\vec{0}$.
Equation \rf{eq:curlcurlb} can be integrated immediately to obtain
\beq
   \curl\vec{b} = 
   ( \vel'\cross\vechat{z})
   - \grad \Psi ,
\eeq
where, by taking the divergence, $\Psi$ solves
\beq
   \label{eq:vortdampsys}
   \left.
      \barr{ll}
         \laplace \Psi = \omega'_z,  &
         \mbox{for{\em\space}}R_1<r<R_2, \\
         \pd{r}\Psi = 0,  &
         \mbox{on{\em\space}}r=R_1,R_2 , 
      \earr
   \right\}
\eeq
and $\omega'_z$ is the $z$-component of vorticity.  
The function $\Psi$ is related to the $z$-component of the 
streamfunction, $\vec{\Phi}$, for three-dimensional incompressible flow 
($\vel=\curl\vec{\Phi}$; $\vec{\omega}=-\laplace\vec{\Phi}$).
The boundary conditions for $\vec{\Phi}$ are determined by the 
no-slip condition.  Here, $\Psi$ satisfies the homogeneous 
Neumann boundary condition that there be no current through the boundaries, 
$\vechat{r}\cdot\curl\vec{b}=0$.  The following analysis may differ
slightly for non-insulating boundaries.
%

The effect of the magnetic field on the
fluid is introduced via the Lorentz force, which can now be written as
\beq
   \label{eq:lorentz}
   \vec{L} = Q \, (\curl\vec{b})\cross\vechat{z} =
    - \, Q \,
    \left[
       u'_r,\, u'_\theta,\, 0\,
    \right]^T
    - Q 
    \left[
       r^{-1} \pd{\theta}\Psi,\, -\pd{r}\Psi,\, 0\,
    \right]^T .
\eeq
The first term on the right hand side is directly proportional and
opposed to any flow across the imposed magnetic field lines.  Axial
flow is unaffected by the Lorentz force.  Therefore it is not surprising
that the flow pattern is found to elongate in this direction as the 
strength of the imposed field increases
(Chandrasekhar 1961).
Taking the curl of \rf{eq:lorentz},
\beq
   \curl \vec{L}  = 
   Q 
   \left(
      \pd{z}\vel \cross \vechat{z} - \omega'_z \, \vechat{z}
   \right)
   -  Q 
   \left(
      \grad\pd{z}\Psi - \omega'_z \, \vechat{z}
   \right) .
\eeq
The $\omega'_z$-terms on the right hand side of \rf{eq:lorentz} 
cancel.
It is well known that components of vorticity perpendicular 
to an imposed field are preferentially damped.
The vorticity of the underlying circular--Couette flow (CCF) points in the
$z$ direction, so CCF is unaffected by the
imposed magnetic field, and the effective viscosity of the fluid is 
unchanged by its presence until the appearance of the Taylor instability.

Taking the dot-product of $\vel'$ with (\ref{eq:gov_limit}{\it a})
and integrating over the volume
gives an energy balance equation for the disturbance.  
The nonlinear term
\beq
   \label{eq:econv}
   \int \vel' \cdot (\vel' \cdot \grad) \vel' \, {\mathrm d}V
   = {\textstyle\frac1{2}} \int \dvgnce (u'^{\,2} \, \vel') \, {\mathrm d}V
\eeq
vanishes due to the boundary conditions.  This term
represents the transfer of energy within the disturbance
to higher modes (smaller length scales) by advection. 
Using the property
\beq
   \int \vel'\cdot\laplace\vel' \, {\mathrm d}V
   = \int \dvgnce (\vel'\cross\vec{\omega}') \, {\mathrm d}V
   - \int \vec{\omega}' \cdot \vec{\omega}' \, {\mathrm d}V ,
\eeq
where similarly the divergence integral vanishes, 
the change in kinetic energy can be written as
\beqarr
   \label{eq:energybal}
   \lefteqn{
      {\textstyle\frac1{2}} \, \pd{t} 
      \int \vel' \cdot \vel' \, {\mathrm d}V
      \,=\, \int u'_r u'_\theta 
      \left(
         \frac1{r} - \pd{r}
      \right)
      \uct \,{\mathrm d}V
   } 
   \nonumber \\
   & & \hspace{15mm}
   - \int \vec{\omega}' \cdot \vec{\omega}' \, {\mathrm d}V
   - \, Q \int
   \left(
      u'^{\,2}_r + u'^{\,2}_\theta - u'_\theta \,\pd{r}\Psi
   \right) 
   \, {\mathrm d}V .
\eeqarr
As the instability that first appears is initially axisymmetric, 
azimuthal derivatives have been ignored in \rf{eq:energybal}.
For a steady-state solution the terms on the right hand side must balance,
hence these three terms can be identified as
\beq
   E_1 - E_2 - E_3 = 0.
\eeq
The first term, $E_1$, is the supply of energy to the disturbance
from radial shear of the circular--Couette flow.  
The second term, $E_2$, is the viscous dissipation and the third, $E_3$,
is the magnetic damping term.

\section{Stability of CCF in a weak magnetic field}

The linear stability of this flow at various radius ratios
(Soundalgekar, Ali \& Takhar 1994), for
co- and counter-rotating cylinders (Chen \& Chang 1998), and at
asymptotically large $Q$ (Chandrasekhar 1961), 
is a well studied topic, so we focus briefly
on the behaviour at small $Q$.  In this and the following sections we 
find the
magnetic field does not need to be large, $Q=\ord(10)$, in order to have 
a significant affect.  By experimental standards this $Q$ is not very 
large; experiments with $Q>10^3$ were performed by \cite{donnelly62}.

The disturbance to the circular--Couette flow can be decomposed
into Fourier modes,
\beq
   \label{eq:decompk}
   \vel'(r,z) = 
   {\textstyle \frac1{\sqrt{2}}}
   \, \vel'_0(r) +
   \sum_{k=1}^\infty 
   \vel'_k(r) \, \cos \alpha kz ,
\eeq
where $2\upi/\alpha$ is the critical wavelength at which the 
fundamental disturbance ($k=1$) first appears.  Due to orthogonality of
the cosine function we have
\beq
   {\textstyle \frac1{2}}
   \int \vel' \cdot \vel' \, {\mathrm d}V 
   = 
   \, \frac{\upi^2}{\alpha} \,
   \sum_k
   \, \int
   \vel'_k \cdot \vel'_k 
   \, r {\mathrm d}r ,
\eeq
which allows us to consider the contribution of each mode to the energy.

For a disturbance to the fundamental mode of  infinitesimal 
amplitude $A_1$, the nonlinear term \rf{eq:econv} will be
very small and modes $k>1$ can be assumed negligible.  
At the critical Reynolds number this disturbance neither grows nor 
decays and can be considered in isolation.


With a fixed outer cylinder the energy, $E_1$, supplied to the fundamental
disturbance is proportional to $\Rey_1$, the driving imposed
by the inner cylinder.  
The inviscid Rayleigh criterion predicts that the flow is unstable
for any rotation of the inner cylinder.
The viscous dissipation, $E_2$, prevents the immediate onset of TVF.
It originates from shears within, and so derivatives of, the disturbance.  
From \rf{eq:energybal} and \rf{eq:decompk}
for the first mode, it is expected that $E_2=\ord(\alpha^2 A_1^2)$.
For a weak magnetic field the magnetic damping, $E_3$, is $\ord(QA_1^2)$, 
linearly proportional to $Q$.   
Provided that the magnetic field is not large enough to affect the
wavenumber of the flow, we expect that the critical driving, 
$\Rey_c$, at which the instability first appears, to be delayed linearly 
with $Q$.  
Further, since $\alpha\approx 3$ for hydrodynamic flows, if $Q=\ord(10)$ 
then $E_2$ and $E_3$ are expected to be of comparable size.  

\begin{figure}
   \bcent
      \epsfig{figure=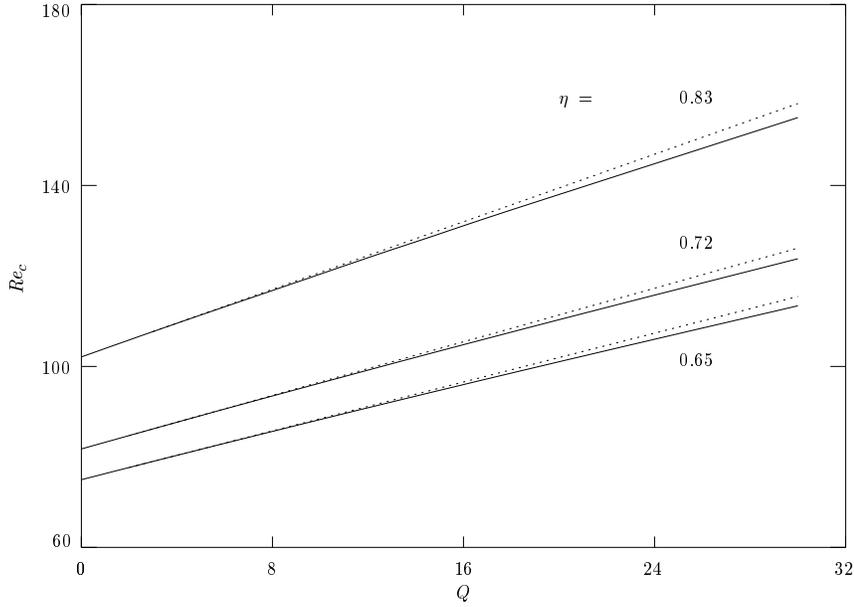, scale=0.71}
      \caption{
         \label{fig:TVFdelay}
         Delay of onset of TVF by an imposed axial field.  Dotted lines
         are linear extrapolations from $Q$ small. 
      }
   \ecent
\end{figure}

Results of calculations at radius ratios $\eta=0.65,\,0.72,\,0.83$
are given in figure \ref{fig:TVFdelay}.
The wavenumber decreases approximately linearly by 15\% 
over the not-so-small range of $Q$ in the figure.
For each $\eta$ the wavenumber $\alpha=3.13, 2.93, 2.70\pm 0.01$ 
at $Q=0, 15, 30$ respectively. 
The delay of 
transition to TVF remains almost linear in $Q$.
In the limit $Q\rightarrow 0$ we find that 
$\Rey_c(Q) / \Rey_c(0) = 1 + \gamma \, Q$ 
where $\gamma\approx 0.0181$ for all $\eta$ in figure \ref{fig:TVFdelay}.  
As no energy is
passed to higher modes for very small $A_1$, the 
structure of the disturbance does not change appreciably with $Q$
or $\eta$.
Therefore the ratio of magnetic to 
viscous dissipation remains the same with different $\eta$.

\section{Nonlinear axisymmetric TVF}

\begin{figure}
   \bcent
      \epsfig{figure=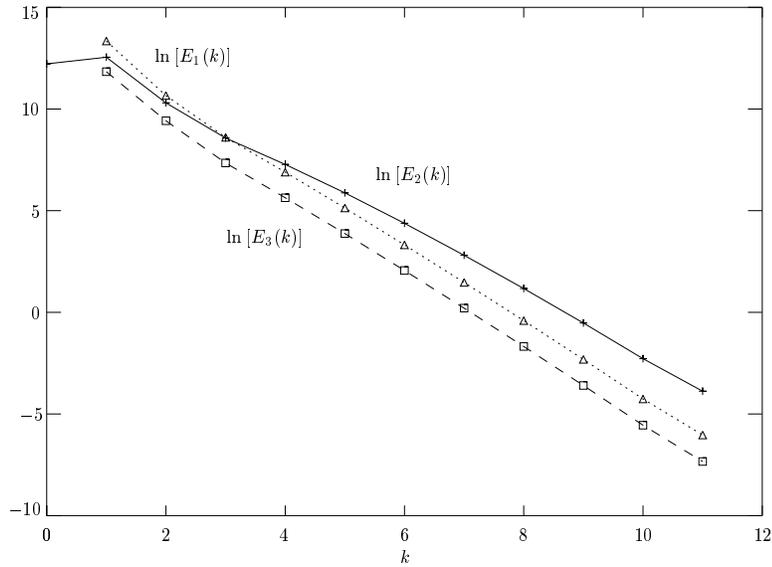, scale=0.71}
   \ecent
      \caption{
         \label{fig:energy_k}
         Typical spectrum of energy terms.
         For the first few modes magnetic (dashed) and 
         viscous (solid) dissipations are 
         comparable.  Viscous dissipation dominates the behaviour
         of higher modes, and $E_2/E_3\propto k^2$.
         ($Q=30$, $\eta=0.65$, $\alpha=2.71$,
         $\Rey_1=1.5\Rey_c$, $\Rey_c=113.4$.)         
      }
\end{figure}

\begin{figure}
   \bcent
      \epsfig{figure=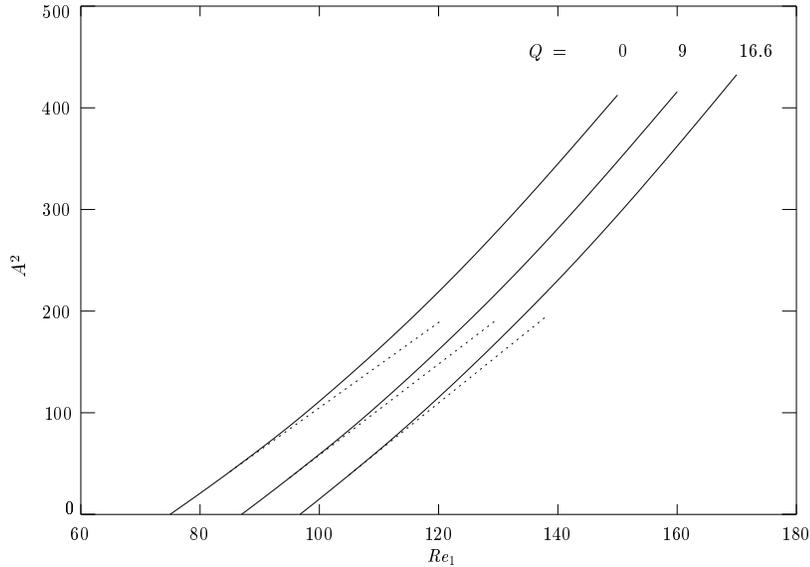, scale=0.71}
      \caption{
         \label{fig:TVFamp}
         Square of maximum axial velocity versus $\Rey_1$ in the presence
         of imposed axial magnetic fields. $\eta=0.65$ .
      }
   \ecent
\end{figure}

In this section we study the nonlinear flows that occur above the first
transition.  
If the inner cylinder is driven past the critical rate, then energy
is quickly transferred to higher modes by the nonlinear advection.
Isolating one of these modes, the dissipative terms are
$E_2=\ord(\alpha^2 k^2 A_k^2)$ and $E_3=\ord(\gamma \, QA_k^2)$.  
Viscous dissipation quickly becomes dominant as energy moves to higher modes.
This can be seen in figure \ref{fig:energy_k}.
However, magnetic damping remains significant on the fundamental mode,
draining energy before it can be passed to higher wavenumbers.  
There the fluid viscosity determines the amplitude of the disturbance.  
The net result can be seen in figure \ref{fig:TVFamp} 
where we plot the amplitude of the disturbance at different values of $Q$.
A relatively wide gap, $\eta=0.65$, was chosen for the calculations,
as the axisymmetric flow is stable to non-axisymmetric perturbations 
well beyond the onset of TVF.
The maximum axial velocity, rather than the radial velocity, was used to
characterise the amplitude of nonlinear TVF, as the outflow becomes
increasingly jet-like as $\Rey_1$ is increased.
\cite{tabeling81} used an amplitude expansion about
the point of criticality in the narrow gap limit
to show that in the weakly nonlinear regime
$A \propto \beta(\Rey_1-\Rey_c)^\frac1{2}$, 
and that $\beta$ does not strongly
depend on $Q$.  From figure \ref{fig:TVFamp}
we see that this approximation is valid well beyond the critical point,
even though the nonlinearity affects the flow pattern.

\cite{kikura99} performed experiments to measure fluid velocities in
nonlinear hydromagnetic flow.
Their imposed field was different from ours, so we cannot make a
direct quantitative comparison.
However, the same order-of-magnitude arguments on the
energy of the harmonics still hold, and the numerical results presented 
in this section are qualitatively the same as their experimental
results. 

\begin{figure}
   \bcent
      \epsfig{figure=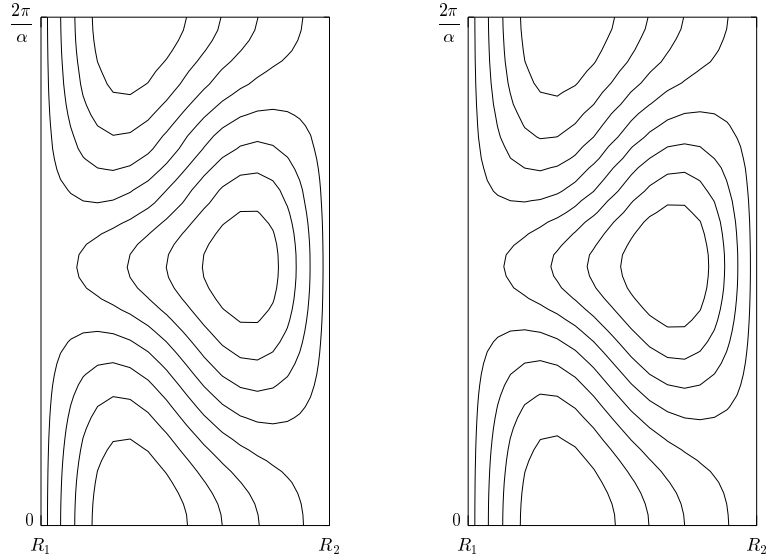, scale=0.71}
      \caption{
         \label{fig:utcont}
         Contours of $u'_\theta$ at $\eta=0.83$ for ({\it a}) $Q=0$,
         $\Rey_1=1.5\Rey_c(Q=0)$, $\alpha=3.13$ and 
         ({\it b}) $Q=20$, $\Rey_1=\Rey_c(Q)+0.5\Rey_c(0)$, 
         $\alpha=2.85$.  
      }
   \ecent
\end{figure}

Since the magnetic damping is effective only on larger length
scales, apart from a change in wavelength of the flow, 
the flow pattern of nonlinear axisymmetric TVF in the presence of 
an imposed axial field is very similar to that in the hydrodynamic 
(non-magnetic) case.  
This is apparent from figure \ref{fig:TVFamp} where the main 
difference is only that the onset of TVF is delayed.
This can also be seen in figure \ref{fig:utcont} where 
$(\Rey_1-\Rey_c(Q))$ is the same for both plots 
(with $\eta=0.83$, in ({\it a}) $Q=0$, $\Rey_1=1.5\Rey_c(0)$, $\alpha=3.13$ 
and ({\it b}) $Q=20$, $\Rey_1=\Rey_c(Q)+0.5\Rey_c(0)$, $\alpha=2.85$).
Although the axial wavelengths differ, 
both are plotted the same size for comparison.
The azimuthal flow is faster than circular--Couette flow $(u'_\theta>0)$
at the outflow regions $(z=\pi/\alpha)$ and slower $(u'_\theta<0)$ at 
the inflow regions $(z=0,\,2\pi/\alpha)$.
The only clear observable difference is that the fast azimuthal
jet-flow region that occurs at the outflow is slightly larger for 
the magnetic flow, 
in addition to the stretched axial wavelength.  
However, figure \ref{fig:utcont}{\it b}
is almost indistinguishable from hydrodynamic flow at the longer
wavelength, $\alpha=2.85$, where $\Rey_1=1.5\Rey_c$ (not plotted).  
The deviation from circular--Couette flow is large; the maximum of
$u'_\theta$ is approximately 35\%
of the maximum of $\uct=\Rey_1$.
The maximum values of $|u'_{r,\theta,z}|$ for all three sets of 
parameters are within 2\%, 
except for $|u'_z|$ in the magnetic flow which is 8\% 
larger.  The Lorentz force has no $z$-component and so 
does not damp axial flow, and therefore the curves in 
figure \ref{fig:TVFamp} are not quite parallel.
Despite only small differences in the flow patterns, 
only the magnetic flow is stable to non-axisymmetric disturbances,
which is the topic of the next section.

\section{The transition to wavy flow}

Finding the preferred mode and transitions between 
fully developed nonlinear wavy-modes is an enormous challenge.  
For the sake of simplicity, we investigate which modes are possible by 
monitoring the growth or decay of infinitesimal non-axisymmetric 
disturbances to the axisymmetric TVF.  The disturbance translates
in the azimuthal direction at some fraction of the rotation rate of the
inner cylinder --- the wavespeed $s$.

\begin{figure}
   \bcent
      \begin{tabular}{c}
         \epsfig{figure=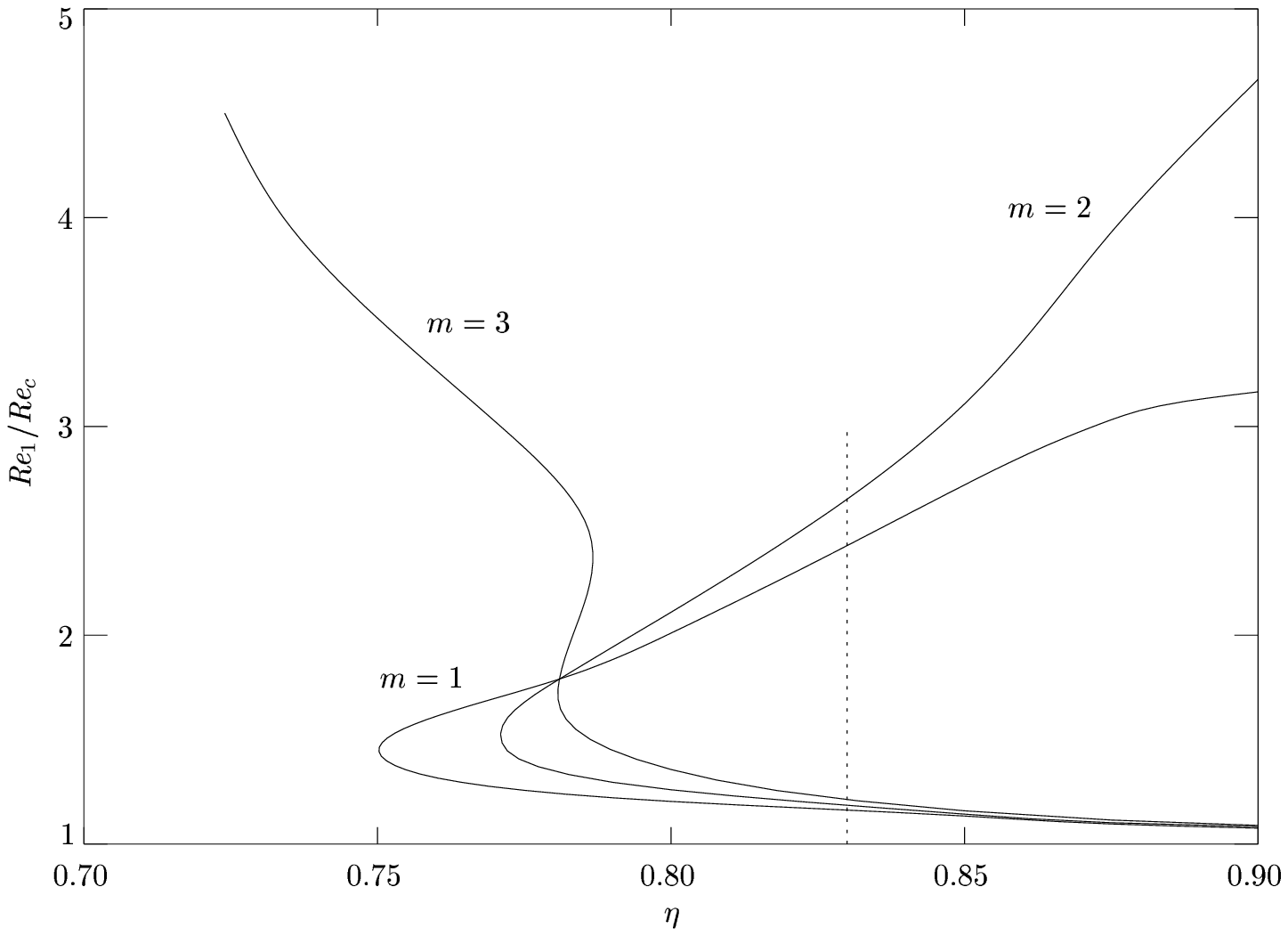, scale=0.71}\\[5pt]
         \epsfig{figure=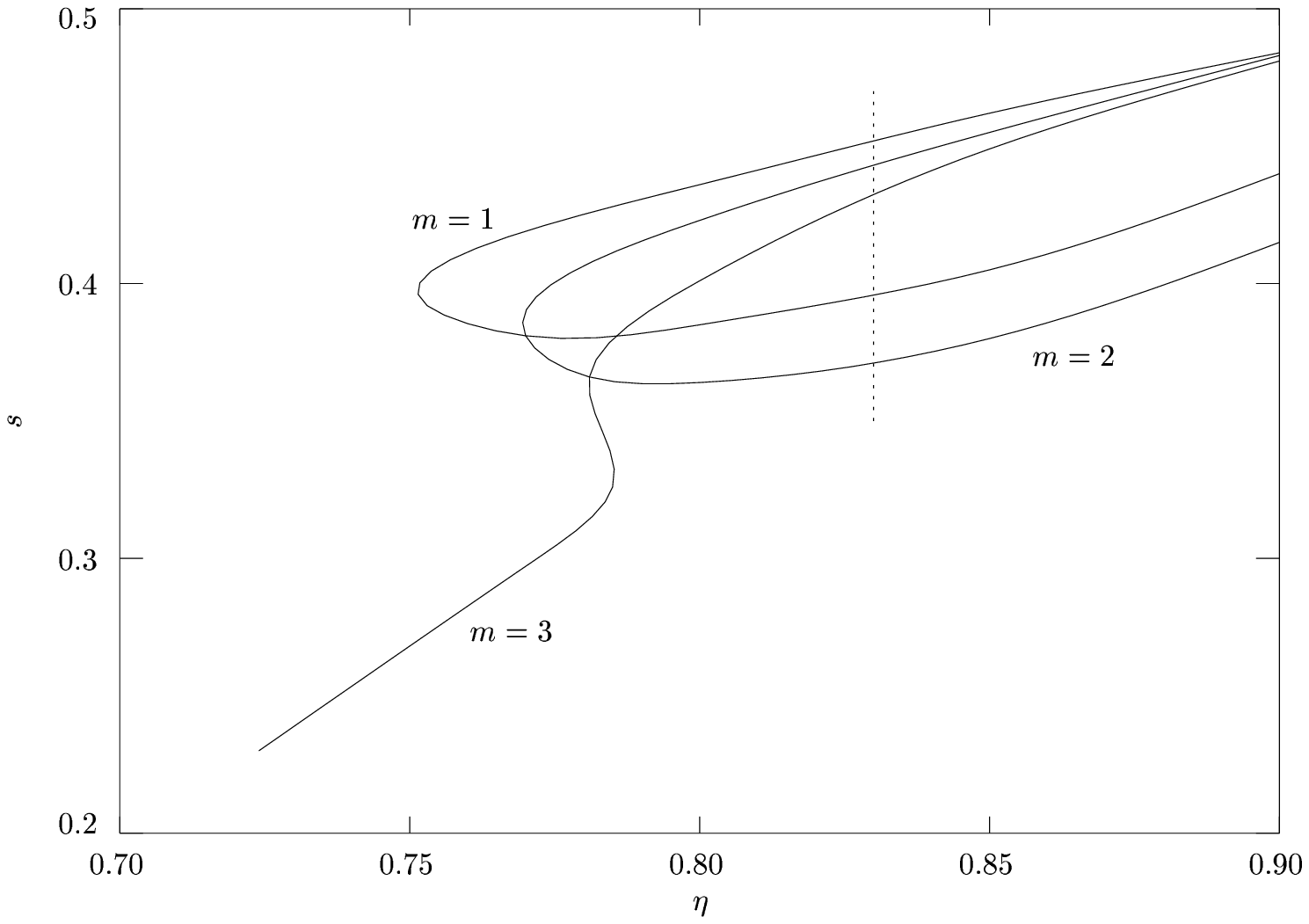, scale=0.71}
      \end{tabular}      
      \caption{
         \label{fig:hydrostab}
         ({\it a}) Stability of axisymmetric TVF to non-axisymmetric 
         perturbations over a range of $\eta$, as
         determined by Jones (1985), $\alpha=3.13$.  
         ({\it b}) Corresponding wavespeeds, $s$, 
         at onset as a fraction of $\Omega_1$. 
         The dotted line is the intersection with the magnetic results
         at $\eta=0.83$ . 
      }
   \ecent
\end{figure}

In the hydrodynamic case, \cite{jones85} calculated stability boundaries
and corresponding wavespeeds by solving the eigenvalue problem.
The stability of axisymmetric flow was found to depend strongly on the 
radius ratio.
Figure \ref{fig:hydrostab} shows the stability boundaries 
and corresponding wavespeeds at onset of the wavy modes.
The stability boundaries are shown
relative to $\Rey_c$ for the onset of TVF $(m=0)$.
For increasing Reynolds numbers and $\eta>0.75$, 
we see that TVF can be destabilised to the $m=1$ wavy mode
(lower part of the boundary in figure \ref{fig:hydrostab}{\it a})
and then be restabilised by a further increase of the Reynolds number
(upper part of the boundary).  Such transition sequences predicted by
Jones' results were verified experimentally by \cite{park84}.

The critical Reynolds number for the disappearance of $m=1,2$ modes 
(upper boundary) increases 
rather rapidly as $\eta$ is increased past about $0.85$.  Therefore
the choice $\eta=0.83$ was taken for our magnetic calculations.  

\begin{figure}
   \bcent
      \begin{tabular}{c}
         \epsfig{figure=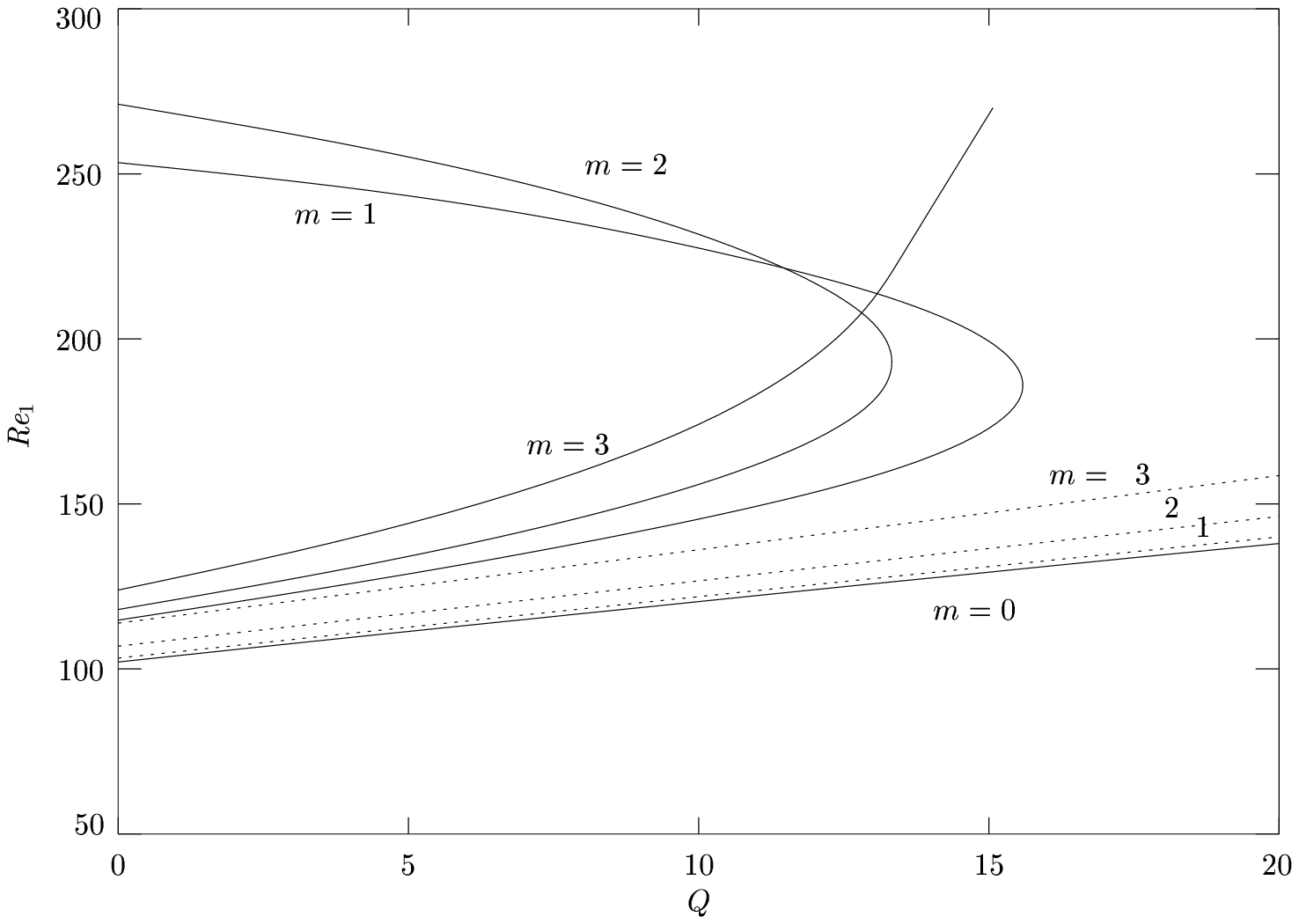, scale=0.71}\\[5pt]
         \epsfig{figure=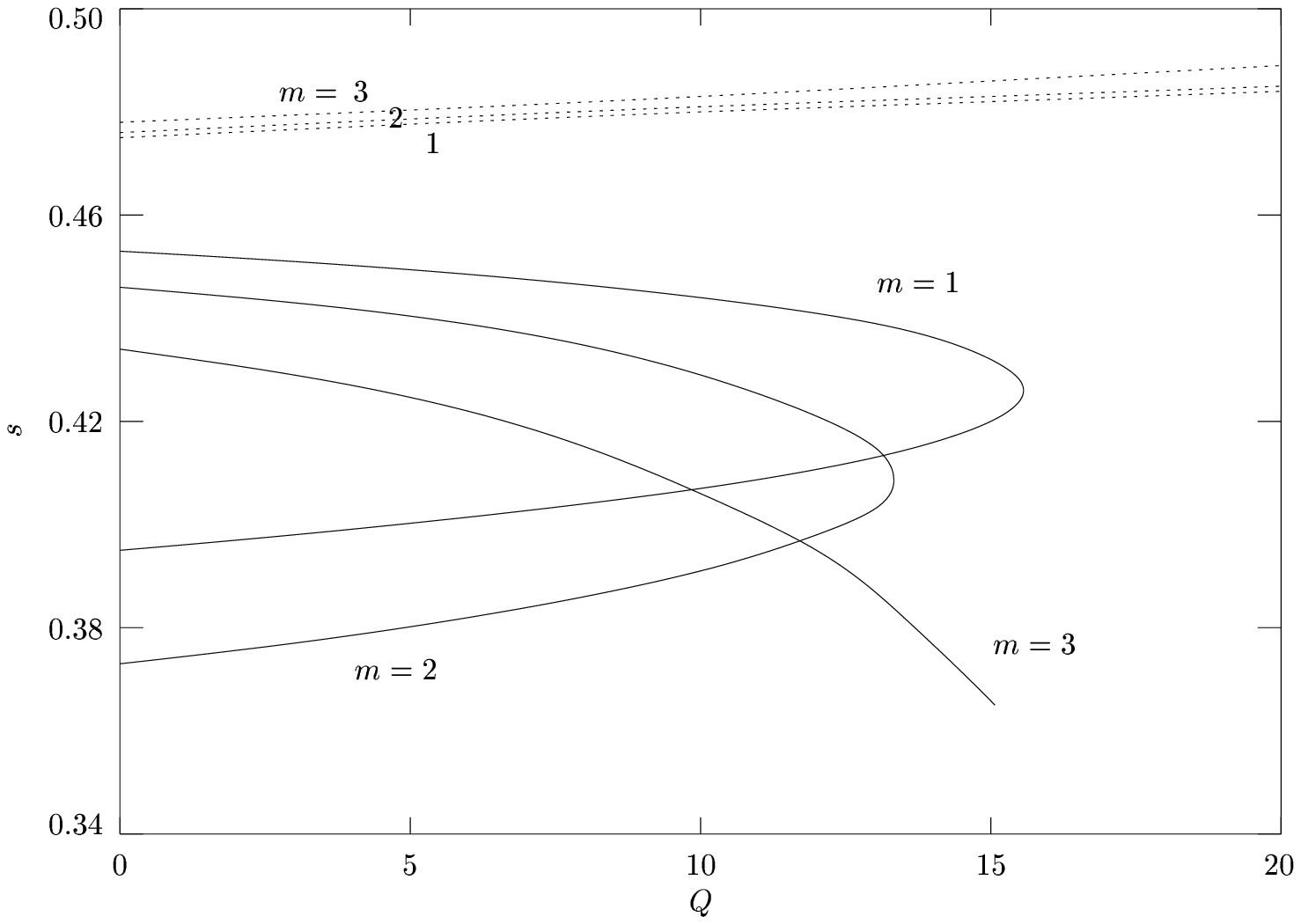, scale=0.71}
      \end{tabular}      
      \caption{
         \label{fig:magstab}
         For $\eta=0.83$, $\alpha = \alpha_c(Q)$
         ({\it a}) stability of hydromagnetic TVF to non-axisymmetric 
         perturbations for increasing magnetic field strength, and
         ({\it b}) corresponding wavespeeds at onset.
         Dotted lines are bifurcations from CCF.          
      }
   \ecent
\end{figure}

The magnetic field, imposed axially, has the effect of increasing
the critical wavelength for onset of TVF.  
\cite{jones85} held $\alpha$ fixed at $3.13$, but this
becomes inappropriate for our magnetic calculations.  
Figure \ref{fig:magstab} shows the stability of the equilibrated TVF 
that appears at the critical wavenumber for the given field strength,
$\alpha(Q)$.
The dotted lines are the stability boundaries for bifurcation from
circular--Couette flow, as calculated by \cite{chen98}.
The difference between the dotted lines and the solid lines confirm
Jones' assertion that it is essential to perturb TVF (which must
be computed numerically) to understand the onset of wavy modes.
The stability boundaries for the onset of wavy modes are sensitive to 
small differences in the parameters, and the magnetic field does 
not need to be strong to produce interesting effects.

The magnetic damping term in \rf{eq:energybal} damps the radial
and azimuthal components of the axisymmetric flow.  
The curvature of the cylinders also has an additional effect on the 
viscous dissipation of these two components of the velocity.
There is a clear similarity 
between increasing the imposed magnetic field strength and decreasing the 
radius ratio.  We do not pursue this relationship
too far as the stability boundaries have
not been fully explained, even for hydrodynamic flows.
However, significant progress on the 
mechanism involved was made by \cite{jones85}. 

\cite{jones85} noted that dissipation due to the radial and axial shear 
within TVF appear to be important factors in the stability of the 
azimuthal jet that appears at the in- and out-flows.
The elongation of Taylor cells over the range of $Q$ in 
figure \ref{fig:magstab} (by around 10\%) 
might be a factor in the suppression of wavy modes.  
\cite{antonijoanpp}
investigated the hydrodynamic stability to wavy modes as a function
of $\alpha$.  
The resulting stability boundaries are qualitatively
similar to figure \ref{fig:magstab}{\it a} for increasing wavelength.
However, 
from the results of \cite{antonijoanpp},
the change in $\alpha$ over the range 
for $Q$ in figure  \ref{fig:magstab}{\it a} is a factor of 2-3 smaller
than that required to suppress  only the $m=1$ mode.  
The enhanced stability must also be attributed to magnetic damping
of the disturbance.

The general trend in figure \ref{fig:hydrostab}{\it b} is for the 
wavespeed, $s$, to decrease with decreasing $\eta$.  
For wider gaps there is more fluid in the slower outer regions and
the mean azimuthal flow speed decreases with increasing gap width,
as does $s$.  
Also, as the Reynolds number is increased the outflow region narrows 
and the inflow region becomes larger than that of the outflow.  
Therefore the bulk of the fluid is in the 
slower inflow region and again the mean azimuthal flow is reduced.
This mean-flow interpretation of the wavespeed is consistent with the 
findings in figure \ref{fig:magstab}{\it b}.  
Here the wavespeed decreases with increasing Reynolds number, as the
outflow still narrows and the inflow broadens, 
but the gap width is fixed and there is no
additional drop in the pattern when the magnetic field strength increases.
Compare figure \ref{fig:hydrostab}{\it b} for 
decreasing $\eta$ with figure \ref{fig:magstab}{\it b}
for increasing $Q$.

\section{Conclusions}

We have investigated magnetic Taylor--Couette flow 
in the nonlinear regime.  In the presence of an imposed axial 
magnetic field the Lorentz force is found to have a significant 
damping affect, but only at larger length scales.  
This is a consequence of the small magnetic Prandtl number limit,
relevant to experiments with liquid metals,
for which the magnetic field is completely defined by the velocity field.

When the imposed field is not too large, the axial wavenumber of the flow
is not greatly affected.  In this weak-field regime the 
stability of circular--Couette flow is enhanced linearly with $Q$,
$\Rey_c(Q) / \Rey_c(0) = 1 + \gamma \, Q$ where $\gamma$ appears to be 
independent of the gap width. 

In the nonlinear regime, magnetic damping affects disturbances at the
fundamental axial wavenumber, but the remaining energy
passed to higher
modes is dissipated mainly by the fluid viscosity.  
This determines the amplitude of the disturbance, which behaves 
approximately like
$\beta(\Rey_1-\Rey_c(Q))^\frac1{2}$
well beyond the critical point, where $\beta$ does not depend 
strongly on $Q$.
This is consistent with the findings of the amplitude expansion by
\cite{tabeling81}.  

Our main finding is that
the magnetic field has its most striking effect on the stability of 
TVF to wavy modes.  A small field is capable of pushing the secondary
instability from only a few percent above the first instability to
several times past the critical Reynolds number for the onset of TVF.
This is similar to the relative stability of TVF to wavy perturbations 
in wide gaps.  As in hydrodynamic flows, the wavespeed decreases with 
increased Reynolds number, but the dependence of the wavespeed on 
the imposed field strength does not appear as strong as the dependence
on the gap width.

The significant enhanced stability observed in the calculations above
occurs at only relatively small imposed field strengths, well within 
experimental range.  
As a secondary bifurcation, the transition to wavy modes is difficult 
to detect accurately via torque measurements.  
A visualisation technique, using ultrasound,
is being developed for opaque fluids by \cite{kikura99}.
This type of transition could serve as a good test for flow visualisation 
in magnetic fluids, where the transitions are accurately defined.
The new technique is of particular interest to those working on dynamo 
experiments.

\begin{acknowledgments}
   The authors wish to thank Prof. Chris Jones for helpful comments
   and suggestions during this work.
\end{acknowledgments}


\end{document}